\documentclass{achemso}
\setkeys{acs}{articletitle = true}
\SectionNumbersOn

\usepackage{chemformula} 
\usepackage[T1]{fontenc} 
\usepackage{multicol}
\usepackage{booktabs}
\usepackage{rotating}
\usepackage{amsmath}\allowdisplaybreaks[1]
\usepackage{braket}
\usepackage{xcolor}
\usepackage{mathrsfs}

\def\dime{hydrogen molecule dimer}

\def\hho{water}

\def\hhoo{(\textit{P})-hydrogen peroxide}

\author{Brendan M. Shumberger}
\author{T. Daniel Crawford}
\email{crawdad@vt.edu}
\affiliation{Department of Chemistry, Virginia Tech, Blacksburg, Virginia, U.S.A.}

\title{Simulation of Vibrational Circular Dichroism Spectra Using Second-Order M{\o}ller-Plesset Perturbation Theory and Configuration Interaction Doubles}

\abbreviations{AATs, CID, MP2}

\begin{document}

\begin{abstract}
We present the first single-reference calculations of the atomic axial tensors (AATs) using
wave-function-based methods including dynamic electron correlation effects using second-order
M{\o}ller-Plesset perturbation theory (MP2) and configuration interaction doubles (CID).  Our implementation
involves computing the overlap of numerical derivatives of the correlated wave functions with respect to both
nuclear displacement coordinates and the external magnetic field.  Out test set included three small
molecules, including the axially chiral \dime\ and \hhoo, and the achiral H$_2$O.  For our molecular test set,
we observed deviations of the AATs for MP2 and CID from that of the Hartree-Fock (HF) method upwards of 49\%,
varying with the choice of basis set.  For \hhoo, electron correlation effects on the VCD rotatory strengths
and corresponding spectra were particularly significant, with maximum deviations of the rotatory strengths of
62\% and 49\% for MP2 and CID, respectively, using our largest basis set.  The inclusion of dynamic electron
correlation to the computation of the AATs can have a significant impact on the resulting rotatory strengths
and VCD spectra.

\end{abstract}

\section{Introduction}

Vibrational circular dichroism (VCD) is one of the unique spectroscopies whose
experimental discovery was preceded by theoretical predictions.  The first
successful attempt to simulate Cotton effects in the infrared region of the
electromagnetic spectrum was reported by Deutsche and Moscowitz in 1968 on a
model helical polymer\cite{Deutsche1968, Deutsche1970}, though the expressions
they derived were not generally applicable.  Experimental measurements of VCD
rotatory strengths followed in 1974 in neat solutions of (\textit{S})-(+)- and
(\textit{R})-(-)-2,2,2-trifluoro-1 -phenylethanol\cite{Holzwarth1974} for
which the only observed vibrational modes were those of the C-H stretching
motions on the chiral carbon.  During this same period, a profusion of
\textit{ad hoc} models for predicting VCD spectra were developed, including
the coupled oscillator model \cite{Holzwarth1972} (1972), the fixed partial
charge model \cite{Schellman1973} (1973), the localized molecular orbital
model \cite{Nafie1977} (1977), the charge flow model \cite{Abbate1981} (1981),
the ring current model \cite{Nafie1983a, Nafie1986} (1983), the atomic polar
tensor model \cite{Freedman1983} and others \cite{Polavarapu1983, Barnett1980,
Barron1979}.  The principal theoretical challenge that spurred these new
models is the fact that the VCD rotatory strength of a given vibrational
transition is zero in the Born-Oppenheimer approximation.  In particular, the
magnetic-dipole vibrational transition moment, whose product with the
corresponding electric-dipole transition moment yields the rotatory strength,
vanishes for adiabatic wave functions.  This unphysical behavior proved highly
nontrivial to overcome through a generalized, first principles approach, and
it was found that these heuristic models exhibited severe limitations that
precluded their widespread applicability\cite{Bursi1990, Stephens1985b}.

In 1983, Nafie and Freedman put forth the vibronic coupling theory
\cite{Nafie1983b}, which superseded the limitations of the BO approximation
\textit{via} introduction of a nuclear perturbation to the adiabatic wave
function.  This approach led to a sum-over-states expression that the authors
were able to reduce to one involving only the ground state wave function using
an average-energy approximation.  In 1985, Stephens introduced the first
fully-ground-state formulation of VCD rotatory strengths \cite{Stephens1985a}.
By introducing first-order perturbations of the ground-state wave function
with respect to the external magnetic field and a nuclear displacement, and
then equating these expressions with the first-order Taylor expansions in the
same variables, he was able to reduce the resulting expression solely to an
overlap of derivatives of the ground-state wave function, known as the atomic
axial tensor (AAT).  Much more recently, a response formalism of VCD was
proposed by Coriani et al.\cite{Coriani2011} which has the advantage that its
implementation into Kohn-Sham density functional theory (KS-DFT) avoided the
need to solve the response functions for the $3N$ geometric displacements,
thereby reducing the computational cost.  To date, Stephens's formulation has
been the most widely implemented approach to calculating the magnetic dipole
vibrational transition moment required to compute VCD.

The first implementation of the AATs of Stephens's VCD formulation was
reported at the Hartree-Fock level by Lowe, Segal, and Stephens using
numerical differentiation of single-determinant wave functions with respect to
nuclear coordinates and external magnetic fields\cite{Stephens1985b,Lowe1986}.
Shortly thereafter Amos, Handy, Jalkanan, and Stephens \cite{Amos1987},
reported the first analytic evaluation of Hartree-Fock AATs based on solutions
to the nuclear and magnetic-field coupled-perturbed Hartree-Fock (CPHF)
equations, validated by the corresponding finite-difference approach.  The
analytic approach is substantially more computationally efficient because it
not only avoids the $6N+6$ calculations for the nuclear displacements and
magnetic-field coordinates, but it also avoids the complex arithmetic required
for finite magnetic fields.

In 1993, Bak \textit{et al.}\cite{Bak1993} reported the anaytic implementation
of AATs at the multiconfigurational self-consistent field (MCSCF) level of
theory.  The authors used an orbital rotation formulation to obtain an
equation for the AATs in terms of density matrices, integrals, derivatives of
molecular orbital coefficients, and derivatives of configuration state
coefficients for which they solved for the coefficient derivatives using the
response formalism developed by Helgaker and J\o rgensen \cite{Helgaker1988}.
In addition, they used gauge-including atomic orbitals (GIAOs) to
circumnavigate the gauge-origin problems that typically plague magnetic-field
dependent properties.  Though they made no direct comparison to experiment,
the authors made note that the correlation effects included by the MCSCF
calculation on NHDT, the isotopomer of ammonia, were significant.  In 1994,
Stephens \textit{et al.}\cite{Stephens94:B3LYP} reported the first application
of M{\o}ller-Plesset (MP) perturbation theory and Kohn-Sham density-functional
theory (KS-DFT) to VCD spectra.  However, only the harmonic force fields were
computed at these levels, while the AATs were still obtained using
Hartree-Fock.  (We note that this was also the first paper to define the B3LYP
exchange-correlation functional, though that was not the focus of the work.)

In 1996, Cheeseman and co-workers\cite{Cheeseman1996} carried out the first
fully analytic DFT-based AAT implementation, including the use of GIAOs to
ensure origin-invariant rotatory strengths.  The results from this
implementation compared well with experiment for trans-2,3 d$_2$-oxirane,
though the authors noted that the accuracy of the DFT results depend on the
density functional adopted, an observation that has been echoed in the the
literature for a number of VCD applications\cite{Nicu08, Autschbach2014,
Grob2023, Cohen2012}.

Here we present the first calculations of VCD AAT rotatory strengths using
dynamically correlated wave-function methods.  In particular, we have
implemented finite-difference gradients of first-order MP and configuration
interaction doubles (CID) wave functions with respect to nuclear coordinates
and external magnetic fields and combined these to obtain the resulting AATs.
We have applied these methods to a number of small-molecule test cases,
including hydrogen peroxide.

\section{Theory}
\subsection{Vibrational Circular Dichroism}

The simulation of VCD spectra requires calculation of the rotatory strength,
$R_{Gg;Gk}$, associated with the $g\rightarrow k$ transition of a given
vibrational mode within the electronic ground state, $G$, is
\begin{equation} \label{RotatoryStrength}
R_{Gg;Gk} = \textrm{Im} \left[ \braket{\Psi_{Gg} | \vec{\mu} | \Psi_{Gk}} \cdot \braket{\Psi_{Gk} | \vec{m} | \Psi_{Gg}} \right],
\end{equation}
where $\vec{\mu}$ and $\vec{m}$ are the electric- and magnetic-dipole
operators, respectively.  In the vibrational harmonic approximation, the
electric-dipole transition moment of the $i$ normal mode is given by the
$\nu=0\rightarrow 1$ transition \cite{Wilson80}
\begin{equation} \label{ElectricDipoleTM}
    \braket{ 0 | \mu_{\beta} | 1 }_i = \left( \frac{\hbar}{2\omega_i} \right)^{1/2}
    \sum_{\lambda\alpha} P_{\alpha\beta}^{\lambda} S_{\lambda\alpha,i},
\end{equation}
where $\omega_i$ is the harmonic angular frequency associated with the normal
mode, $P_{\alpha\beta}^{\lambda}$ are the atomic polar tensors (APTs), and
$S_{\lambda\alpha,i}$ is the normal coordinate transformation matrix from
Cartesian nuclear displacements to mass-weighted normal mode displacements.
In this notation, $\beta$ denotes a particular Cartesian direction of the
external electric field, $\lambda$ indexes the nuclei, and $\alpha$ is a
Cartesian coordinate of the $\lambda$-th nucleus.  The APTs are typically
computed using the electrical harmonic approximation, \textit{i.e.}, as
derivatives of the molecular dipole moment with respect to nuclear
displacements,
\begin{equation} \label{APT}
    P_{\alpha\beta}^\lambda = \left(\frac{\partial\braket{\Psi_G|\mu_{\beta}|\Psi_G}}{\partial R_{\lambda\alpha}}\right)_{R_{\lambda\alpha} = R^0_{\lambda\alpha}},
\end{equation}
where $\braket{\Psi_G|\mu_{\beta}|\Psi_G}$ is the electric-dipole moment
expectation value in the electronic ground state, and $R^0_{\lambda\alpha}$
denotes the equilibrium/reference geometry.

Similarly, the magnetic-dipole transition moment is expressed as,
\begin{equation} \label{MagneticDipoleTM}
    \braket{ 0 | m_{\beta} | 1 }_i = \left( 2 \hbar^3 \omega_i \right)^{1/2}
    \sum_{\lambda\alpha} M_{\alpha\beta}^{\lambda} S_{\lambda\alpha,i},
\end{equation}
where $M_{\alpha\beta}^{\lambda}$ is the AAT, $\lambda$ and $\alpha$ have the
same meaning as in Eq.~\eqref{APT}, and $\beta$ denotes a particular Cartesian
direction of the external magnetic field.  The AAT can be separated into its
nuclear and electronic components as
\begin{equation} \label{AAT}
    M_{\alpha\beta}^{\lambda} = J_{\alpha\beta}^{\lambda} + I_{\alpha\beta}^{\lambda}.
\end{equation}
The nuclear contribution is given by
\begin{equation} \label{NuclearAAT}
    J_{\alpha\beta}^{\lambda} = \frac{i}{4 \hbar} \sum_{\gamma} \epsilon_{\alpha\beta\gamma} R_{\lambda\gamma}^0 Z_{\lambda} e,
\end{equation}
where $\epsilon_{\alpha\beta\gamma}$ is the three dimensional Levi-Civita
tensor, $R_{\lambda\gamma}^0$ is the $\gamma$-th equilibrium Cartesian
coordinate of the $\lambda$-th nucleus, and $Z_{\lambda} e$ is the charge of
the $\lambda$-th nucleus. 

The primary focus of the present work is the electronic contribution to the
AAT, which, in Stephens's formulation is\cite{Stephens1985a}
\begin{equation} \label{ElectronicAAT}
    I_{\alpha\beta}^{\lambda} = \left\langle
    \left( \frac{\partial \Psi_G(R)}{\partial R_{\lambda\alpha}} \right)_{R_{\lambda\alpha} = R^0_{\lambda\alpha}} \bigg|
    \left( \frac{\partial \Psi_G(R_0, H_{\beta})}{\partial H_{\beta}} \right)_{H_\beta=0} \right\rangle.
\end{equation}
Thus, the challenge is to compute the overlap of two derivatives of the
ground-state wave function: one with respect to nuclear displacements,
$R_{\lambda\alpha}$, and one with respect to the external magnetic field,
$H_{\beta}$, both evaluated at the equilibrium geometry and at zero-field.

\subsection{Electronic AATs within the MP2 and CID Wavefunction Approximations}
In both MP2 and CID the electronic wave functions are of the form
\begin{equation} \label{Wavefunctions}
    \left| \Psi_\textrm{corr} \right\rangle = ( 1 + \hat{T}_2 ) \left| \Phi_{0} \right\rangle
\end{equation}
where $\Psi_{\textrm{corr}}$ is the correlated wave function (MP2 or CID), $\hat{T}_2$
is the double-excitation operator, and $\Phi_{0}$ is the reference
Hartree-Fock (HF) wave function.  In spin-orbital notation, the $\hat{T}_2$
operator is given by 
\begin{equation} \label{Amplitudes}
\hat{T}_2 = \frac{1}{4} \sum_{ijab} t_{ij}^{ab} a_a^{\dagger} a_b^{\dagger} a_j a_i,
\end{equation}
with $i$, $j$, $k$, and $l$ to represent occupied orbitals and $a$, $b$, $c$,
and $d$ to virtual orbitals.  The cluster amplitudes, $t_{ij}^{ab}$ in
Eqs.~\eqref{Wavefunctions} and \eqref{Amplitudes} are obtained from the
first-order wave function equation for MP2, 
\begin{equation}
\label{MP2_amps}
\braket{\Phi_{ij}^{ab}|\hat{H}|\Phi_0} + \braket{\Phi_{ij}^{ab}|\left(\hat{H}^{(0)} - E^{(0)}\right) \hat{T}_2|\Phi_0} = 0,
\end{equation}
where $|\Phi_{ij}^{ab}\rangle$ is a doubly excited determinant, $\hat{H}$ is
the electronic Hamiltonian, $\hat{H}^{(0)}$ is the Fock operator, and
$E^{(0)}$ is the sum of the occupied orbital energies.  For CID, the
amplitudes are obtained from the projection of the CI Schr\"odinger equation
onto the doubly excitated determinants,
\begin{equation}
\label{CID_amps}
\braket{\Phi_{ij}^{ab}|\hat{H} \left(1 + \hat{T}_2\right)|\Phi_0} - E_\textrm{CID} t_{ij}^{ab} = 0.
\end{equation}
Stephens's construction of the AAT in Eq.~\eqref{ElectronicAAT} requires that
the ground-state wave function is fully normalized, but the above expressions
assume intermediate normalization.  Thus, after the amplitudes have been
computed, they must be renormalized such that 
\begin{equation}
\braket{\Psi_\textrm{corr} | \Psi_\textrm{corr}} = 1.
\end{equation}

Incorporation of the normalized form of Eq.~\eqref{Wavefunctions} into 
Eq.~\eqref{ElectronicAAT} yields four terms,
\begin{align}
I_{\lambda\alpha}^{\beta} &= 
|c_0|^2 \Braket{\frac{\partial \Phi_0}{\partial R_{\lambda\alpha}} | \frac{\partial \Phi_0}{\partial H_\beta}} + 
c_0 \Braket{\frac{\partial \Phi_0}{\partial R_{\lambda\alpha}} | \frac{\partial}{\partial H_\beta} \left(\hat{T}_2 \Phi_0\right)} + \nonumber \\
& c_0 \Braket{\frac{\partial}{\partial R_{\lambda\alpha}} \left(\hat{T}_2 \Phi_0\right) | \frac{\partial \Phi_0}{\partial H_\beta}} +
\Braket{\frac{\partial}{\partial R_{\lambda\alpha}} \left(\hat{T}_2 \Phi_0\right) | \frac{\partial}{\partial H_\beta} \left(\hat{T}_2 \Phi_0\right)},
    \label{SOAATs}
\end{align}
where $c_0$ denotes the coefficient of $\Phi_0$ in the fully normalized wave
function.  As shown by Pearce,\cite{Pearce2021} the second and third terms on
the right-hand side are equal in magnitude, but opposite in sign, and thus
exactly cancel.  Following Stephens and 
co-workers,\cite{Stephens1985b,Lowe1986,Amos1987} we have computed the above
derivatives using a finite-difference approach in which the HF and MP2/CID
wave function is computed after the displacement of a given nuclear
coordinate or by the addition of a weak magnetic field in a chosen direction
to the Hamiltonian.  For each combination of displaced wave functions, their
overlap is obtained by multiplying the overlap of the corresponding
determinants with the associated $\hat{T}_2$ amplitudes and/or $c_0$
coefficients.  However, because the displaced wave functions are in different
MO bases, the overlaps between Slater determinants are computed by taking the
determinant of the MO-basis overlap matrix of the orbitals comprising the bra
and ket determinants.  For excited determinants, this matrix is obtained by
swapping the appropriate rows (bra) and/or columns (ket) corresponding to the
occupied and virtual spin-orbitals.

As pointed out by Stephens \textit{et al.},\cite{Stephens1985b,Lowe1986} one
must carefully account for the phases of the displaced wave functions in order
to obtain physically correct values of the wave function overlaps in the
finite-difference scheme.  For the nuclear coordinate displacements, the HF
and correlated wave functions remain real, and so the phase factors are
associated only with the signs of the individual HF MOs.  For magnetic-field
displacements, the wave functions are complex, and thus the MO coefficients
and cluster amplitudes are also complex.  Following Stephens,\cite{Lowe1986}
the presence of a magnetic-field perturbation in the Hamiltonian yields a
complex phase factor between two normalized MOs,
\begin{equation}
\phi_p'(\vec{R}^0, H_\beta) = e^{i\theta}\phi_p(\vec{R}^0, H_\beta),
\end{equation}
where $\phi_p(\vec{R}^0, H_\beta)$ is the MO that reduces to
$\phi_p(\vec{R}^0)$ in the absence of the field.  For weak fields, the overlap
of $\phi_p'(\vec{R}^0, H_\beta)$ with the zero-field MO is
\begin{equation} \label{Phase}
\Braket{\phi_p(\vec{R}^{0}) | \phi_p'(\vec{R}^{0},H_{\beta})}
    = \Braket{\phi_p(\vec{R}^{0}) | \phi_p(\vec{R}^{0},H_{\beta})} e^{i\theta} 
    = N e^{i\theta},
\end{equation}
where $N$ is the (real) normalization constant of
$\phi_p'(\vec{R}^{0},H_{\beta})$ through second order in the field.  Squaring
this overlap yields the square of $N$,
\begin{equation}
\left|\Braket{\phi_p(\vec{R}^{0}) | \phi_p'(\vec{R}^{0},H_{\beta})}\right|^2 = N^2,
\end{equation}
which allows us to obtain the value of $e^{i\theta}$ from the overlap in
Eq.~\eqref{Phase} and correct the phase of $\phi_p'(\vec{R}^0, H_\beta)$.
(Note that this same approach also serves to correct the phases on the MOs
for real perturbations, in which case the phase factor simplifies to $\pm1$.)
Once the phases of the MOs have been defined, the phases of the amplitudes are
also defined because of the projection-based approach we use in 
Eqs.~\eqref{MP2_amps} and \eqref{CID_amps}.

\section{Computational Details}

We have implemented the finite-difference scheme described above in an open-source Python code,
MagPy\cite{magpy}, which uses the form given in Eq.~\eqref{SOAATs} to compute the AAT.  Given that MP2 and CID
AATs have not yet been reported in the literature, we have tested and validated this code in several ways: (1)
we have developed a second, independent Python implementation that splits the derivatives of the cluster
amplitudes and the Slater determinants into separate finite-difference calculations; (2) we have derived and
coded both spin-orbital and spin-adapted (spatial orbital) forms for both MP2 and CID AATs; (3) we have
compared our (non-GIAO) Hartree-Fock AATs to those produced by the DALTON code\cite{Dalton} for all test
cases; (4) we have computed all four contributions appearing in Eq.~\eqref{SOAATs} to confirm that the second
and third terms cancel.  Both of these implementations use the Psi4 quantum chemistry package\cite{PSI4} to provide the
necessary Hamiltonian, electric-/magnetic-dipole, and mixed-basis-set overlap integrals.

To analyze the impact of dynamic electron correlation effects on the electronic AATs, we have selected three
molecular test cases, \dime, \hho, and \hhoo, as well as numerous basis sets: STO-3G, 6-31G, 6-31G(d),
cc-pVDZ, and cc-pVTZ, depending on the size of the molecule.\cite{Hehre1969, Ditchfield1971, Hehre1972,
Hariharan1973, Dunning1989, Pritchard2019, Feller1996, Schuchardt2007} The \dime\ is included because it is
the smallest possible chiral system, which has been used in previous studies as representative of the
beginning of a helical hydrogen system for benchmarking calculations of optical rotation.\cite{D'Cunha2023}
Our chosen (non-optimized) structure of the \dime\ has an intramolecular H$-$H distance of 0.75 \AA, an
intermolecular H$-$H distance on the helical backbone of 1.0 \AA, and a $+60^{\circ}$ dihedral angle.  The
geometries for \hho\ and \hhoo\ were optimized at the MP2/cc-pVDZ level of theory using the CFOUR quantum
chemistry package.\cite{CFOUR} [Geometries of all three test molecules are given in the Supporting Information
(SI).] The $1s$ core orbitals of the oxygen atoms were frozen in all MP2 and CID calculations on \hho\ and
\hhoo.

We report rotatory strengths and corresponding VCD spectra only for \hhoo, because it is the only chiral
compound in our test set for which the the equilibrium geometry remains chiral (for basis sets that include
polarization functions).  Inherent performance limitations of Python as well as the cost of evaluating the
overlap of two doubly excited determinants in different bases in Eq.~\eqref{SOAATs} precluded application of
the code to larger molecules.  In order to focus on the impact of dynamic electron correlation effects on the
spectra, we used a common geometry and harmonic force field (both computed at the MP2/cc-pVDZ level of theory)
for all basis sets.  In addition, because we do not include GIAOs in our implementation, the VCD rotatory
strengths we report are origin dependent. As such, keeping the geometry fixed reduces the impact of
re-optimizing the molecular structure for each basis set.

\section{Results and Discussion}

\subsection{Atomic Axial Tensors}

In order to provide a simple quantitative comparison of these tensors, we report correlation minimum/maximum
percent changes, which are obtained by dividing each element of a given correlated AAT by its Hartree-Fock
counterpart, subtracting $1.0$ from each element and converting to a percentage.  This is intended to provide
an estimate of magnitude of electron correlation effects on the AAT elements.  The corresponding impact on VCD
rotary strengths will be discussed in the next section.

The AATs computed for the \dime\ at the HF, MP2, and CID levels of theory are given in Table \ref{dimer
cc-pVDZ} using cc-pVDZ basis set. (The results from the STO-3G, 6-31G, and cc-pVTZ basis sets are provided in
the SI.) The minimum/maximum percentages of the HF AATs for MP2 are 98.5/99.0\% for
STO-3G, 97.7/102.4\% for 6-31G, 95.6/101.6\% for cc-pVDZ, and 97.6/106.2\% for cc-pVTZ.  For CID, the
minimum/maximum percentages are 96.2/98.1\% for STO-3G, 95.2/107.5\% for 6-31G, 83.5/105.6\% for cc-pVDZ, and
92.6/121.3\% for cc-pVTZ.  As is clear, the gap between the minimum and maximum percentages increases for both
MP2 and CID as the basis set size grows.  Additionally, for each basis set, the difference between the maximum
and minimum values is always greater for CID over MP2. This difference reaches a maximum of 28.8\% for the
\dime\ using CID with the cc-pVTZ basis (see Table S7 of the SI).

\begin{table}[h!]
    \scriptsize
        \caption{Electronic HF, MP2, and CID AATs (a.u.) for the \dime \ using the cc-pVDZ basis.}
    \label{dimer cc-pVDZ}
    \begin{tabular}{@{}lccccccccccc}
    \toprule
    & \multicolumn{3}{c}{HF} & \hspace{1pt} & \multicolumn{3}{c}{MP2} & \hspace{1pt} & \multicolumn{3}{c}{CID} \\
    \cmidrule{2-4} \cmidrule{6-8} \cmidrule{10-12}
    & $B_x$ & $B_y$ & $B_z$ & & $B_x$ & $B_y$ & $B_z$ & & $B_x$ & $B_y$ & $B_z$ \\
    \midrule
    H$_{1x}$ & -0.078893 & -0.008996 &  0.364298 & & -0.078875 & -0.008600 &  0.362977 & & -0.078636 & -0.007509 &  0.362176 \\
    H$_{1y}$ &  0.031734 &  0.009958 &  0.073418 & &  0.031340 &  0.009951 &  0.073174 & &  0.030302 &  0.009932 &  0.073471 \\
    H$_{1z}$ & -0.529172 & -0.138045 &  0.085432 & & -0.527501 & -0.137850 &  0.085348 & & -0.525801 & -0.138038 &  0.085000 \\
    H$_{2x}$ & -0.053675 & -0.025007 &  0.342479 & & -0.053706 & -0.025413 &  0.341287 & & -0.053566 & -0.026407 &  0.340754 \\
    H$_{2y}$ & -0.024080 & -0.008875 &  0.109311 & & -0.023638 & -0.008864 &  0.108857 & & -0.022607 & -0.008877 &  0.108197 \\
    H$_{2z}$ & -0.523197 & -0.123777 &  0.078075 & & -0.521843 & -0.123201 &  0.078051 & & -0.520677 & -0.122274 &  0.077854 \\
    H$_{3x}$ & -0.053675 & -0.025007 & -0.342479 & & -0.053706 & -0.025413 & -0.341287 & & -0.053566 & -0.026407 & -0.340754 \\
    H$_{3y}$ & -0.024080 & -0.008875 & -0.109311 & & -0.023638 & -0.008864 & -0.108857 & & -0.022607 & -0.008877 & -0.108197 \\
    H$_{3z}$ &  0.523197 &  0.123777 &  0.078075 & &  0.521843 &  0.123201 &  0.078051 & &  0.520677 &  0.122274 &  0.077854 \\
    H$_{4x}$ & -0.078893 & -0.008996 & -0.364298 & & -0.078875 & -0.008600 & -0.362977 & & -0.078636 & -0.007509 & -0.362176 \\
    H$_{4y}$ &  0.031734 &  0.009958 & -0.073418 & &  0.031340 &  0.009951 & -0.073174 & &  0.030302 &  0.009932 & -0.073471 \\
    H$_{4z}$ &  0.529172 &  0.138045 &  0.085432 & &  0.527501 &  0.137850 &  0.085348 & &  0.525801 &  0.138038 &  0.085000 \\
    \bottomrule
    \end{tabular}
\end{table}

The \hho\ molecule provides a reasonable test case for investigating the effect of electron correlation on the
AATs, just as it has for a range of other properties.  Although the molecule is achiral, it contains multiple
non-zero AAT elements for comparison between methods.  AATs for \hho\ are given in Table \ref{h2o cc-pVDZ} for
HF, MP2, and CID, again using the cc-pVDZ basis set.  (Results obtained for the remaining basis sets are
provided in the SI.)  The $C_{2v}$ symmetry, with the molecule lying the $yz$-plane and the $C_2$ axis along
the $z$-axis, limits the number of non-zero tensor components.  In particular, only elements for which the
direct product of the irreducible representations (irreps) of the $R_{\lambda\alpha}$ displacement and the
$H_\beta$ field component contains the totally symmetric irrep can be non-vanishing.  The $H_\beta$ terms
transform as rotations about the three Cartesian axes, $H_x\rightarrow B_1$, $H_y\rightarrow B_2$, and
$H_z\rightarrow A_2$, but only the bare coordinate displacements of the oxygen atom transform as $C_{2v}$
irreps, $O_x\rightarrow B_1$, $O_y\rightarrow B_2$, and $O_z\rightarrow A_1$.  Symmetry-adapted linear
combinations of the coordinate displacements of the hydrogen atoms transform as irreps, \textit{e.g.}, $H_{1x}
- H_{2x}\rightarrow A_2$.  These properties govern the pattern of vanishing values of the H$_2$O AATs in Table
\ref{h2o cc-pVDZ}, as well as those of the other basis sets reported in the SI.

The changes in the maximum and minimum correlated percentages are not as significant as that of the \dime.
The minimum/maximum percentages associated with MP2 are 99.5/101.7\%, 99.3/101.5\%, 100.1/101.5\%, and
100.7/102.1\% for STO-3G, 6-31G, 6-31G(d), and cc-pVDZ, respectively.  For CID the minimum/maximum percentages
are 99.3/104.6\%, 99.4/102.8\%, 101.0/102.6\%, and 101.4/104.0\% for the same basis-set ordering.  The largest
gap between the minimum and maximum percentages occurs for CID/STO-3G at only 5.3\%.  Furthermore, the trend
in differences between the minimum and maximum percentages are not nearly as consistent as the \dime\ test
case.  For MP2 the maximum/minimum percentage gap is 6-31G > STO-3G > 6-31G(d) > cc-pVDZ where the gap size
ranges from 1.4\% to 2.2\%, whereas for CID, the gap size trend as STO-3G > 6-31G > cc-pVDZ > 6-31G(d) with
the minimum value being 1.6\%.  For all the basis sets, the gap size is larger for CID compared to MP2.

\begin{table}[h!]
    \scriptsize
        \caption{Electronic HF, MP2, and CID AATs (a.u.) for \hho \ using the cc-pVDZ basis.}
    \label{h2o cc-pVDZ}
    \begin{tabular}{@{}lccccccccccc}
    \toprule
    & \multicolumn{3}{c}{HF} & \hspace{1pt} & \multicolumn{3}{c}{MP2} & \hspace{1pt} & \multicolumn{3}{c}{CID} \\
    \cmidrule{2-4} \cmidrule{6-8} \cmidrule{10-12}
    & $B_x$ & $B_y$ & $B_z$ & & $B_x$ & $B_y$ & $B_z$ & & $B_x$ & $B_y$ & $B_z$ \\
    \midrule
    O$_{1x}$ & -0.000000 & -0.046076 &  0.000000 & & -0.000000 & -0.047037 &  0.000000 & & -0.000000 & -0.047906 &  0.000000 \\
    O$_{1y}$ &  0.105707 & -0.000000 &  0.000000 & &  0.107082 & -0.000000 &  0.000000 & &  0.107556 & -0.000000 &  0.000000 \\
    O$_{1z}$ &  0.000000 &  0.000000 &  0.000000 & & -0.000000 &  0.000000 &  0.000000 & & -0.000000 &  0.000000 &  0.000000 \\
    H$_{2x}$ &  0.000000 &  0.069789 & -0.101645 & &  0.000000 &  0.070693 & -0.102471 & &  0.000000 &  0.071165 & -0.103160 \\
    H$_{2y}$ & -0.069867 &  0.000000 & -0.000000 & & -0.070357 &  0.000000 & -0.000000 & & -0.070850 &  0.000000 & -0.000000 \\
    H$_{2z}$ &  0.111684 & -0.000000 & -0.000000 & &  0.112634 & -0.000000 & -0.000000 & &  0.113396 & -0.000000 & -0.000000 \\
    H$_{3x}$ & -0.000000 &  0.069789 &  0.101645 & & -0.000000 &  0.070693 &  0.102471 & & -0.000000 &  0.071165 &  0.103160 \\
    H$_{3y}$ & -0.069867 & -0.000000 &  0.000000 & & -0.070357 & -0.000000 &  0.000000 & & -0.070850 & -0.000000 &  0.000000 \\
    H$_{3z}$ & -0.111684 &  0.000000 & -0.000000 & & -0.112634 &  0.000000 & -0.000000 & & -0.113396 &  0.000000 & -0.000000 \\
    \bottomrule
    \end{tabular}
\end{table}

The third test case, \hhoo, provides an optimized chiral system for which both AATs and VCD spectra may be
examined.  The AATs for \hhoo\ using the cc-pVDZ basis are provided in Table \ref{h2o2 cc-pVDZ}.  (Again, the
results from the remaining basis sets are provided in the SI.) The minimum/maximum percentages for the MP2
method are 97.2/107.6\%, 89.0/104.4\%, 84.0/148.9, and 92.5/104.6 using the STO-3G, 6-31G, 6-31G(d) and
cc-pVDZ basis sets, respectively.  For CID, we observe similar changes to MP2 where the minimum/maximum
percentages are 93.6/105.8\%, 91.0/102.7\%, 86.5/116.4\%, and 94.2/103.9\% for the same basis sets,
respectively.  In contrast to both \dime\ and \hho, the largest difference between the minimum and maximum
values (64.9\%) was observed using the MP2 method with the 6-31G(d) basis set.  However, similar to \hho,
there is no clear trend between the basis set size and the minimum/maximum gap.  For MP2, the gap size varies 
as 6-31G(d) > 6-31G > cc-pVDZ > STO-3G and for CID, 6-31G(d) > STO-3G > 6-31G > cc-pVDZ.

\begin{table}[h!]
    \scriptsize
        \caption{Electronic HF, MP2, and CID AATs (a.u.) for \hhoo \ using the cc-pVDZ basis.}
    \label{h2o2 cc-pVDZ}
    \begin{tabular}{@{}lccccccccccc}
    \toprule
    & \multicolumn{3}{c}{HF} & \hspace{1pt} & \multicolumn{3}{c}{MP2} & \hspace{1pt} & \multicolumn{3}{c}{CID} \\
    \cmidrule{2-4} \cmidrule{6-8} \cmidrule{10-12}
    & $B_x$ & $B_y$ & $B_z$ & & $B_x$ & $B_y$ & $B_z$ & & $B_x$ & $B_y$ & $B_z$ \\
    \midrule
    H$_{1x}$ & 0.004090 & -0.032185 &  0.092323 & &  0.004015 & -0.031457 &  0.092030 & &  0.004057 & -0.032279 &  0.092993 \\
    H$_{1y}$ & 0.056218 & -0.089054 &  0.350998 & &  0.056866 & -0.093126 &  0.357087 & &  0.057078 & -0.092099 &  0.355837 \\
    H$_{1z}$ &-0.093657 & -0.274700 &  0.085311 & & -0.094740 & -0.277656 &  0.088809 & & -0.095085 & -0.277008 &  0.087994 \\
    H$_{2x}$ & 0.004090 & -0.032185 & -0.092323 & &  0.004015 & -0.031457 & -0.092030 & &  0.004057 & -0.032279 & -0.092993 \\
    H$_{2y}$ & 0.056218 & -0.089054 & -0.350998 & &  0.056866 & -0.093126 & -0.357087 & &  0.057078 & -0.092099 & -0.355837 \\
    H$_{2z}$ & 0.093657 &  0.274700 &  0.085311 & &  0.094740 &  0.277656 &  0.088809 & &  0.095085 &  0.277008 &  0.087994 \\
    O$_{3x}$ &-0.008638 &  0.065415 & -0.109005 & & -0.008641 &  0.064745 & -0.105808 & & -0.008649 &  0.065425 & -0.106124 \\
    O$_{3y}$ &-0.014022 & -0.046288 &  2.120337 & & -0.014337 & -0.042809 &  2.113230 & & -0.014573 & -0.043610 &  2.114382 \\
    O$_{3z}$ & 0.063390 & -2.049988 &  0.058282 & &  0.064332 & -2.046831 &  0.055502 & &  0.064796 & -2.048017 &  0.056045 \\
    O$_{4x}$ &-0.008638 &  0.065415 &  0.109005 & & -0.008641 &  0.064745 &  0.105808 & & -0.008649 &  0.065425 &  0.106124 \\
    O$_{4y}$ &-0.014022 & -0.046288 & -2.120337 & & -0.014337 & -0.042809 & -2.113230 & & -0.014573 & -0.043610 & -2.114382 \\
    O$_{4z}$ &-0.063390 &  2.049988 &  0.058282 & & -0.064332 &  2.046831 &  0.055502 & & -0.064796 &  2.048017 &  0.056045 \\
    \bottomrule
    \end{tabular}
\end{table}

\subsection{Rotatory Strengths and Vibrational Circular Dichroism Spectra}

The inclusion of dynamic electron correlation on the VCD spectra of \hhoo\ produces multiple effects,
depending on the choice of basis set.  The harmonic vibrational frequencies, IR intensities, and rotatory
strengths of \hhoo\ for HF, MP2, and CID for the STO-3G, 6-31G, 6-31G(d), and cc-pVDZ basis sets are given in
Tables \ref{h2o2 vcd STO-3G} - \ref{h2o2 vcd cc-pVDZ}.  We note again that the same MP2/cc-pVDZ geometry and
harmonic force constants were used for all levels of theory and basis sets in order to more clearly separate
the impact of dynamic electron correlation on the VCD rotatory strengths from other effects, such as the
optimized bond lengths and bond angles.

The most intense transition across all basis sets corresponds to the dihedral bending motion occurring at
338.53 cm$^{-1}$.  However, while this transition is the most prominent in the VCD spectrum, its intensity,
when introducing correlation effects, only deviates from that of HF by 2$-$7\% across all basis sets.  For the
STO-3G basis, only the symmetric H$-$O$-$O bending transition at 1443.26 cm$^{-1}$ exhibits smaller deviations
from HF than the dihedral bend.  For the 6-31G and cc-pVDZ sets, on the other hand, this transition shifts by
22$-$34\%, whereas 6-31G(d) by only ~8\% for both MP2 and CID.

The next two most intense VCD bands correspond to the symmetric and anti-symmetric hydrogen stretching modes
at 3812.87 cm$^{-1}$ and 3810.34 cm$^{-1}$.  The former exhibits relatively strong correlation shifts relative
to HF with STO-3G and 6-31G deviating between 32$-$57\% and 6-31G(d) and cc-pVDZ deviating between 16$-$23\%.
Similarly, the anti-symmetric O$-$H stretch changes significantly --- by 69\% -- from HF at the CID/STO-3G
level, a significant variation for this intense vibrational band.  Using the STO-3G basis, the greatest
deviation for MP2 is observed at 920.51 cm$^{-1}$ for the oxygen stretching motion which changes by 57\%,
though this is also the weakest vibrational transition.  For 6-31G, 6-31G(d), and cc-pVDZ, the transition most
affected was the antisymmetric hydrogen bending motions observed at 1306.96 cm$^{-1}$ for which deviations
ranged between 29\% and 90\%.

\begin{figure}
\centering
\includegraphics[width=\textwidth]{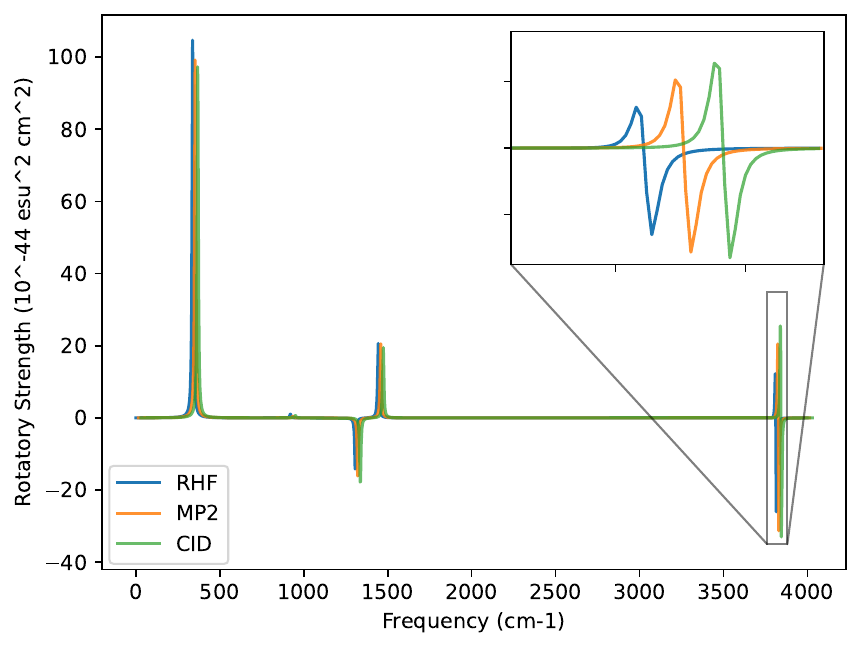}
\caption{VCD spectra of \hhoo\ using a common geometry and Hessian. The geometry and Hessian were
    computed using the cc-pVDZ  basis at the MP2 level. The APTs and AATs were computed using the
    STO-3G basis set. The full-width half-maximum was set to 16 cm$^{-1}$. For readability, the
    MP2 and CID spectra were shifted by 15 cm$^{-1}$ and 30 cm$^{-1}$, respectively.}
\label{h2o2 STO-3G VCD}
\end{figure}

The VCD spectra generated from these data are presented in Figures \ref{h2o2 STO-3G VCD} - \ref{h2o2 cc-pVDZ
VCD}, and we observe four notable features of these spectra across the various basis sets.  First, the STO-3G
basis exhibits changes in the signs of the rotatory strengths relative to the larger basis sets for all
transitions above 1000 cm$^{-1}$, even though the geometry and harmonic force constants used are identical in
all cases.  (We have confirmed this behavior at the HF level using the DALTON code.) Second, going from the
6-31G to the 6-31G(d) basis introduces a sign change for the transition at 920.51 cm$^{-1}$, though, again,
this is the weakest transition in the spectrum.  Third, all three basis sets larger STO-3G tend to produce
magnitudes of the rotatory strengths in the order of HF > CID > MP2 where the optimization of the CI
coefficients corrects the MP2 transition intensity back towards that of the HF value.  Finally, the symmetric
hydrogen stretch at 3812.87 cm$^{-1}$ becomes slightly washed out by the anti-symmetric stretch at 3810.34
cm$^{-1}$ as the basis set size increases and correlation effects are included.  This is a direct result of
our choice of full-width at half-max, the near degeneracy of the two modes, and the difference in absolute
rotatory strength between the symmetric and antisymmetric hydrogen stretching motions.

\begin{table}[h!]
    \caption{Frequencies, IR intensities, and rotatory strengths for \hhoo \ using the STO-3G basis.
        Quantities were obtained using a common MP2/cc-pVDZ geometry and Hessian.}
    \label{h2o2 vcd STO-3G}
    \begin{tabular}{@{}lcccccccc}
    \toprule
    Frequency & \hspace{1pt} & \multicolumn{3}{c}{IR Intensity} & \hspace{1pt} & \multicolumn{3}{c}{Rotatory Strength} \\
    (cm$^{-1}$) & \hspace{1pt} & \multicolumn{3}{c}{(km/mol)} & \hspace{1pt} & \multicolumn{3}{c}{($10^{-44}$ esu$^2$ cm$^2$)} \\
    \cmidrule{0-0} \cmidrule{3-5} \cmidrule{7-9}
    & & HF & MP2 & CID & & HF & MP2 & CID \\
    \midrule
    3812.87  & &  16.026 &  27.795 &  33.997 & & -55.443 &  -73.296 & -81.774 \\
    3810.34  & &  31.963 &  60.122 &  82.195 & &  46.473 &   67.168 &  78.475 \\
    1443.26  & &   1.213 &   1.198 &   1.113 & &  21.365 &   21.260 &  20.191 \\
    1306.96  & &  47.401 &  45.301 &  48.395 & & -14.985 &  -17.013 & -18.822 \\
     920.51  & &   0.068 &   0.013 &   0.029 & &   1.071 &    0.460 &   0.660 \\
     338.53  & & 134.248 & 118.958 & 113.545 & & 106.476 &  100.907 &  98.960 \\
    \bottomrule
    \end{tabular}
\end{table}

\begin{table}[h!]
    \caption{Frequencies, IR intensities, and rotatory strengths for \hhoo \ using the 6-31G basis.
    Quantities were obtained using a common MP2/cc-pVDZ geometry and Hessian.}
    \label{h2o2 vcd 6-31G}
    \begin{tabular}{@{}lcccccccc}
    \toprule
    Frequency & \hspace{1pt} & \multicolumn{3}{c}{IR Intensity} & \hspace{1pt} & \multicolumn{3}{c}{Rotatory Strength} \\
    (cm$^{-1}$) & \hspace{1pt} & \multicolumn{3}{c}{(km/mol)} & \hspace{1pt} & \multicolumn{3}{c}{($10^{-44}$ esu$^2$ cm$^2$)} \\
    \cmidrule{0-0} \cmidrule{3-5} \cmidrule{7-9}
    & & HF & MP2 & CID & & HF & MP2 & CID \\
    \midrule
    \midrule
    3812.87  & &  15.191 &   2.533 &   3.309 & &  53.349 &  22.911  &  26.009 \\
    3810.34  & &  72.759 &  20.018 &  24.491 & & -63.938 & -28.768  & -31.378 \\
    1443.26  & &   0.490 &   0.256 &   0.299 & & -17.797 & -12.826  & -13.823 \\
    1306.96  & & 118.048 & 139.448 & 134.127 & &   8.545 &   0.817  &   1.713 \\
     920.51  & &   2.594 &   1.248 &   1.665 & &   3.749 &   2.470  &   2.799 \\
     338.53  & & 320.104 & 291.536 & 294.062 & & 261.392 & 250.415  & 251.641 \\
    \bottomrule
    \end{tabular}
\end{table}

\begin{table}[h!]
    \caption{Frequencies, IR intensities, and rotatory strengths for \hhoo \ using the 6-31G(d) basis.
    Quantities were obtained using a common MP2/cc-pVDZ geometry and Hessian.}
    \label{h2o2 vcd 6-31Gd}
    \begin{tabular}{@{}lcccccccc}
    \toprule
    Frequency & \hspace{1pt} & \multicolumn{3}{c}{IR Intensity} & \hspace{1pt} & \multicolumn{3}{c}{Rotatory Strength} \\
    (cm$^{-1}$) & \hspace{1pt} & \multicolumn{3}{c}{(km/mol)} & \hspace{1pt} & \multicolumn{3}{c}{($10^{-44}$ esu$^2$ cm$^2$)} \\
    \cmidrule{0-0} \cmidrule{3-5} \cmidrule{7-9}
    & & HF & MP2 & CID & & HF & MP2 & CID \\
    \midrule
    3812.87  & &  23.053 &  11.213 &  12.794 & &  24.458 &  19.814 &  20.477 \\
    3810.34  & &  94.143 &  50.167 &  56.910 & & -36.618 & -29.131 & -29.883 \\
    1443.26  & &   0.478 &   0.411 &   0.406 & & -20.796 & -19.249 & -19.080 \\
    1306.96  & & 120.017 & 136.651 & 131.984 & &  18.794 &  12.799 &  13.311 \\
     920.51  & &   2.745 &   1.690 &   2.044 & &  -2.835 &  -2.254 &  -2.473 \\
     338.53  & & 236.828 & 222.557 & 224.529 & & 222.566 & 216.812 & 217.700 \\
    \bottomrule
    \end{tabular}
\end{table}

\begin{table}[h!]
    \caption{Frequencies, IR intensities, and rotatory strengths for \hhoo \ using the cc-pVDZ basis.
    Quantities were obtained using a common MP2/cc-pVDZ geometry and Hessian.}
    \label{h2o2 vcd cc-pVDZ}
    \begin{tabular}{@{}lcccccccc}
    \toprule
    Frequency & \hspace{1pt} & \multicolumn{3}{c}{IR Intensity} & \hspace{1pt} & \multicolumn{3}{c}{Rotatory Strength} \\
    (cm$^{-1}$) & \hspace{1pt} & \multicolumn{3}{c}{(km/mol)} & \hspace{1pt} & \multicolumn{3}{c}{($10^{-44}$ esu$^2$ cm$^2$)} \\
    \cmidrule{0-0} \cmidrule{3-5} \cmidrule{7-9}
    & & HF & MP2 & CID & & HF & MP2 & CID \\
    \midrule
    3812.87  & &  30.781 &  13.757 &  16.506 & &  32.728 &  25.002 &  26.586 \\
    3810.34  & & 117.644 &  57.086 &  67.839 & & -50.910 & -38.440 & -40.476 \\
    1443.26  & &   0.246 &   0.106 &   0.132 & & -11.812 &  -7.731 &  -8.590 \\
    1306.96  & & 105.238 & 114.319 & 110.611 & &  11.921 &   4.499 &   6.029 \\
     920.51  & &   2.456 &   1.292 &   1.683 & &  -3.257 &  -2.396 &  -2.735 \\
     338.53  & & 217.281 & 192.586 & 196.575 & & 152.732 & 143.478 & 144.888 \\
    \bottomrule
    \end{tabular}
\end{table}

\clearpage

\begin{figure}
\centering
\includegraphics[width=\textwidth]{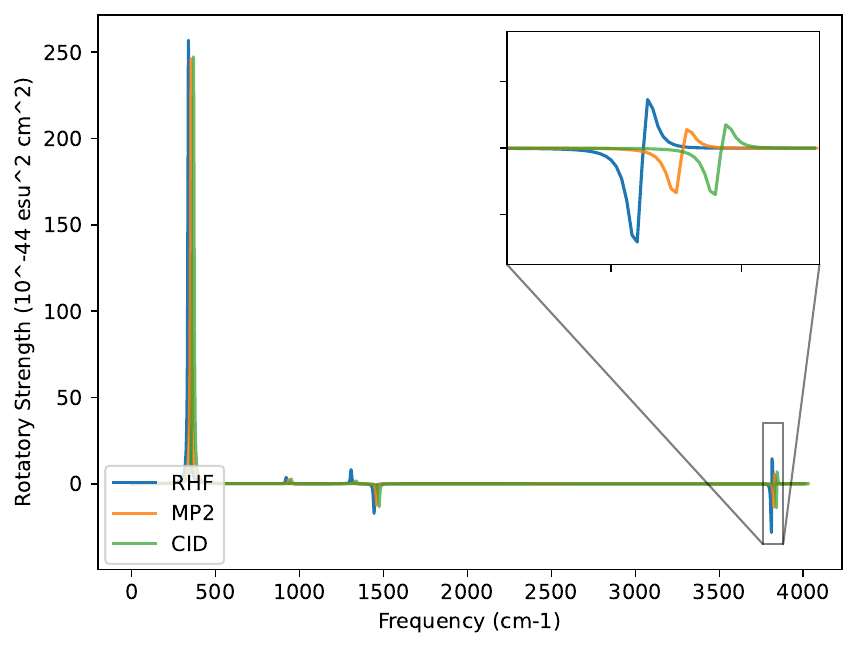}
\caption{VCD spectra of \hhoo\ using a common geometry and Hessian. The geometry and Hessian were
    computed using the cc-pVDZ  basis at the MP2 level. The APTs and AATs were computed using the
    6-31G basis set. The full-width half-maximum was set to 16 cm$^{-1}$. For readability, the
    MP2 and CID spectra were shifted by 15 cm$^{-1}$ and 30 cm$^{-1}$, respectively.}
\label{h2o2 6-31G VCD}
\end{figure}
\clearpage

\begin{figure}
\centering
\includegraphics[width=\textwidth]{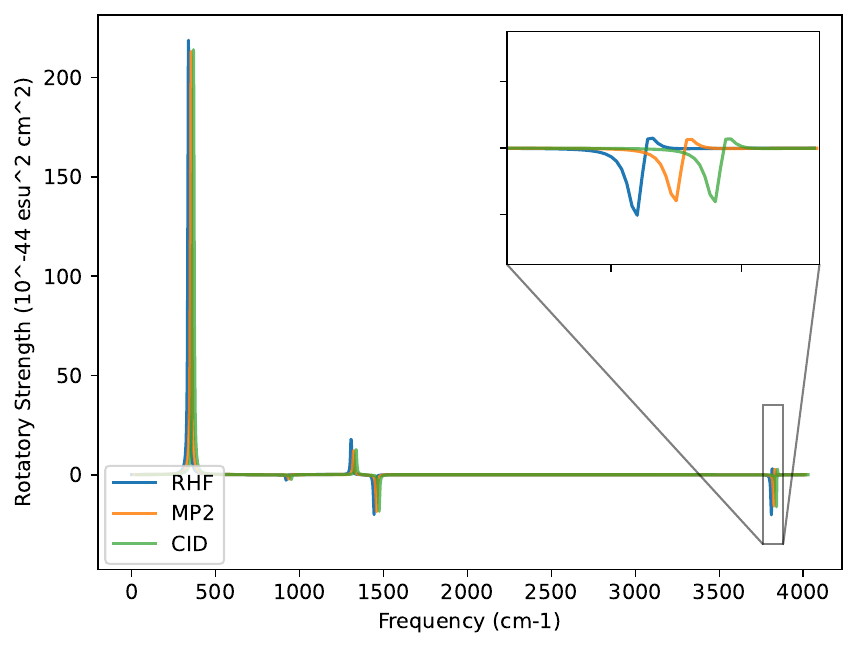}
\caption{VCD spectra of \hhoo\ using a common geometry and Hessian. The geometry and Hessian were
    computed using the cc-pVDZ  basis at the MP2 level. The APTs and AATs were computed using the
    6-31G(d) basis set. The full-width half-maximum was set to 16 cm$^{-1}$. For readability, the
    MP2 and CID spectra were shifted by 15 cm$^{-1}$ and 30 cm$^{-1}$, respectively.}
\label{h2o2 6-31Gd VCD}
\end{figure}
\clearpage

\begin{figure}
\centering
\includegraphics[width=\textwidth]{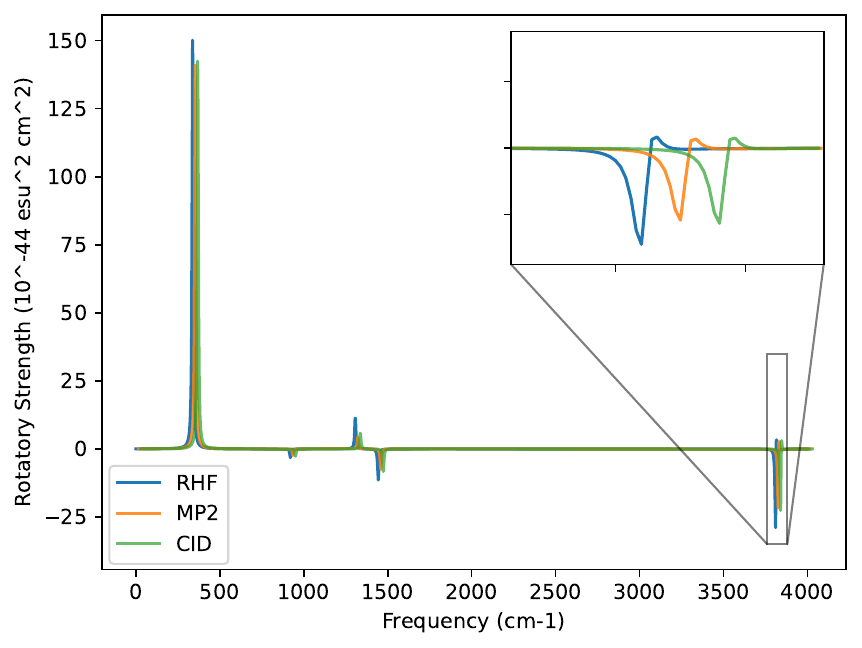}
\caption{VCD spectra of \hhoo\ using a common geometry and Hessian. The geometry and Hessian were
    computed using the cc-pVDZ  basis at the MP2 level. The APTs and AATs were computed using the 
    cc-pVDZ basis set. The full-width half-maximum was set to 16 cm$^{-1}$. For readability, the 
    MP2 and CID spectra were shifted by 15 cm$^{-1}$ and 30 cm$^{-1}$, respectively.}
\label{h2o2 cc-pVDZ VCD}
\end{figure}
\clearpage

\section{Conclusion}

We have reported the first simulations of VCD spectroscopy including dynamic electron correlation using wave
function methods, specifically the MP2 and CID levels of theory. Our implementation relies on a
finite-difference scheme to obtain the Hessian, APTs, and AATs in the Cartesian coordinate basis.  While the
Hessian and APTs can be formulated as second derivatives of the energy, the AATs are formulated as overlaps
between wave function derivatives due to the fact that the required electronic contributions to the
magnetic-dipole transition moments are unphysically zero within the Born-Oppenheimer approximation.
Subsequent transformation into the normal coordinate basis yields the vibrational frequencies, infrared
intensities, and rotatory strengths required for simulating the absorption and VCD spectra.

We benchmarked our implementation using three small test cases including \dime, \hho, and \hhoo.  The effects
of correlation on the AATs are much more significant in the two chiral molecules than that of the achiral
system (H$_2$O) reaching a maximum deviation from the uncorrelated method of 21\% (CID/cc-pVTZ) for \dime\ and
49\% (MP2/6-31G(d)) for \hhoo\, while \hho\ only reaching a maximum deviation of 5\% (CID/STO-3G).  These
effects appear concomitantly in the VCD spectra of \hhoo\ where the rotatory strength yields maximum
deviations of 90\% (MP2/6-31G).  We note that five of the six transition intensities are of the incorrect sign
for the small STO-3G basis set when compared to that of the cc-pVDZ set for every level of theory.
Additionally, for the 6-31G, 6-31G(d), cc-pVDZ, we noted that optimization of the CI coefficients tends to
correct the MP2 transition intensities back in the direction of the HF intensities with the maximum correction
being approximately 13\% (cc-pVDZ).

This work provides an avenue to benchmark future implementations of VCD using the MP2, CID, and higher levels
of theory using analytic gradient methods.  Due to the high computational scaling associated with computing
determinants of nonorthogonal basis functions, future work will be directed at developing such analytic
schemes.

\section{Supporting Information} \label{si}

Atomic coordinates, AATs, and rotatory strengths of the test molecules are available.

\section{Acknowledgements} \label{ack}

TDC was supported by the U.S.\ National Science Foundation via grant CHE-2154753 and BMS by grant DMR-1933525.
The authors are grateful to Advanced Research Computing at Virginia Tech for providing computational resources
that have contributed to the results reported within the paper.

\bibliography{references}

\end{document}